# Evolution of clusters in the energetic heavy ion bombarded amorphous graphite

**M. N. Akhtar[1], Bashir Ahmad[1] and Shoaib Ahmad[1,2]**

[1]Pakistan Institute of Nuclear Science & Technology, P. O. Nilore, Islamabad, Pakistan.

[2]National Centre for Physics, QAU Campus, Shahdara Valley Road, Islamabad 44000, Pakistan

Email: sahmad.ncp@gmail.com

## Abstract

A broad range of carbon clusters have been generated by a novel technique of energetic heavy ion bombardment of amorphous graphite. The evolution of clusters and their subsequent fragmentation under continuing ion bombardment is revealed by detecting the charged clusters in the energy spectra of the Direct Recoils emitted as a result of collision between the incident ions and surface where the newly formed clusters are residing. The techniques of producing and detecting carbon clusters from $C_2$ to ones $>C_{100}$ have been indigenously developed at PINSTECH.

Ever since the discovery of Buckminsterfullerene [1,2], carbon clusters have extensively been produced and studied with various techniques including high temperature and high pressure production of soot [3,4], laser beam ablation of graphite [5,6,7] and by electronic sputtering of organic films by MeV ions [8,9]. These techniques have their respective merits and advantages. Whereas, there has been an emphasis in these techniques to selectively produce the magic clusters, some authors [10,11,12] have indicated that much heavier clusters than $C_{60}$ can also be produced.

In the present communication we report a series of experiments that has been conducted to investigate mechanisms of production as well as monitoring of clusters recoiling from amorphous graphite surface under heavy ion bombardment. The bombarding ions initiate cluster forming mechanisms and can also eject the carbon complexes by imparting them sufficient energy as Direct Recoils. Measurement of successive energy spectra of these recoils provide us with the signatures of the dynamics of heavy ion induced cluster formation in amorphous graphite. Direct Recoils emitted from the target consist of the monatomic, diatomic, triatomic and higher cluster forms of pure carbon. These recoil after elastic binary collisions with the mono-energetic ionic projectiles and carry with them energies $E_r$ which are function of target to projectile mass ratio ($m_2 / m_1$), angle of recoil $\theta_r$ and the bombarding energy $E_0$ and is given as

$$E_r = 4.\{(m_1.m_2) / (m_1+m_2)^2\}.E_0 \, . \cos^2 \theta_r.$$

Energy analysis is performed with a 90-degree electrostatic energy analyzer. Measurements of successive energy spectra of these recoils provide us with the signatures of the dynamics of heavy ion-induced cluster formation on the surface of the irradiated graphite.

The experiments were planned in such a way that heavy ions deposit considerable amounts of energy (~100-150 eV/Å) in a region determined by ion's range. By choosing ion beam incidence angle with surface normal $\alpha = 80°$, a wedge like ion irradiated region is produced of not more than 30 atomic layers beneath the surface and with the ion penetration depth ~ 500-760 Å at 100 keV for $Xe^+$ and $Kr^+$, respectively. The incident ionic flux ensures that in addition to the generation of intense collision cascades, maximum randomization of the target constituents occurs. This may lead to the simulation of laser beam ablated carbon targets [5] which have been instrumental in the generation of fullerenes.

This technique has three essential features which make it useful for the study of formation of carbon clusters especially the fullerenes;

a) almost all existing bonds between carbon atoms along the ion path are broken on the time scale ~ $10^{-15}$ sec,

b) the mobile secondary knock-ons being in ionized/excited states continue to produce energetically the most favorable carbon complexes on a time scale typical of the cooling down of the resulting collision cascades (~$10^{-13}$ sec), and

c) the Direct Recoils that originate in binary collisions between incident ions and the surface constituents carry with them the information that characterizes the dynamics of cluster formation under ion bombardment.

Amorphous graphite was chosen to ensure a structure-less carbon medium that would undergo cyclic sequences of bond breaking and re-bonding of carbon atoms to form complexes. The Direct Recoils produced as a byproduct of ion-target interactions are detected at large recoil angles. The only constraint on choosing large recoil angles is that these recoils with energies ~ few tens to just over a 100 eV should be clearly distinct from the collision cascade sputtering contributions i.e., particles emitted with few eV and yet have sufficient velocities (typically ~ $10^5$ [cm/s]) to be detected by a channelton electron multiplier [13,14]. We have shown that successive energy spectra of DRs emitted after collisions with the target constituents reveal the dynamics of cluster formation as well as the fragmentation processes resulting from the incident ion - cluster collisions.

Our previous observation [15] has been that under $Ar^+$ and $Kr^+$ bombardment a large fraction of carbon monomers ($C_1^{n+}$) recoil at $\theta_r$ ~ 80° from graphite surface with high probability of being in multiply charged states i.e., n≥1. There were indications, however, that multi-mers ($C_m^+, m \geq 2$ ) were also present in the spectra but definitive identification required variation in the ion species as well as the recoil angle $\theta_r$ .The present energy spectra of direct recoils were taken at values of recoil angle $\theta_r$ = 87.8°. Such a large recoil angle was chosen due to its correspondingly higher LSS differential cross sections dσ [16]. Since $d\sigma \propto E_r^{-1}$, where $E_r$ is the energy of target particles in a direct recoil with the projectile. $E_r$ reduces for large $\theta_r$ thereby considerably enhancing dσ. There has been at least an order of magnitude increase in the measured cross sections in going from our earlier $\theta_r$ = 79.5° to the present value 87.8°.

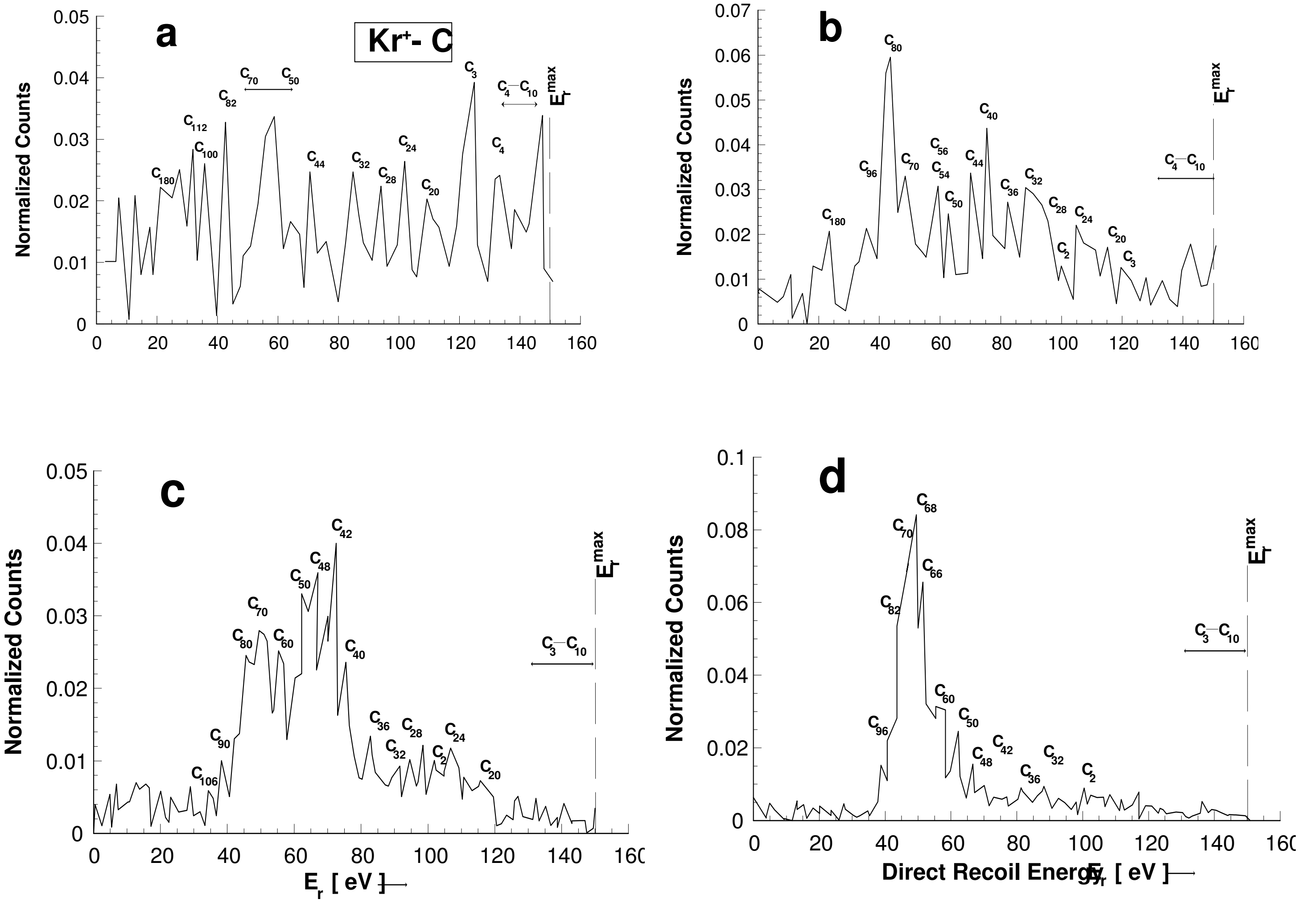

**Figure (1)**. Shows the Direct Recoil energy spectra from $Kr^+$-C at 100 kV of Carbon clusters $C_m$ are seen recoiling with characteristic energies $E_r$. Fig (a) to Fig (d) present the dynamic behaviour of the formation and gradual fragmentation of various clusters. $E_r^{max}$ is the maximum transferable energy to a cluster in a direct recoil.

Figure (1) shows the energy spectra of direct recoils from 100 keV $Kr^+$ ion bombarded graphite at recoil angle $\theta_r = 87.8°$. Four consecutive spectra are shown with 4 μA beam incident at grazing incidence $\alpha$ ~ 80°. The first spectrum (1-a) is taken after a dose of $2.5x10^{13}$ ions. The entire fullerene range is present with a broad peak around $C_{60}^{+}$, higher m as well as the lighter ones including m<36 and the linear/chain structural combinations ( m= 1 to ~10). The next spectrum (1-b) has $C^+_{82}$ as the dominant species while other cluster especially the fragments $C^+_{m-2}$ resulting from $C_m \rightarrow C_{m-2} + C_2$ e.g., $C^+_{80}$, $C^+_{56}$, $C^+_{44}$, $C^+_{40}$ are conspicuous by their relative abundance. The lower order fullerenes have increased their share of the total yield. $C^+_{70}$ and $C^+_{50}$ are present but $C^+_{60}$ is not significantly present. The gradual building up of the $C^+_{50}$ and its fragments ($C^+_{48}$, $C^+_{46}$, $C^+_{42}$, $C^+_{40}$) can be seen from fig (1-c). A well defined peak for $C^+_{60}$ which compares well with $C^+_{70}$, $C^+_{80}$, $C^+_{82}$ and $C^+_{50}$. The smaller clusters are present but their total yield is much smaller than that of the higher (i.e., > $C^+_{50}$) fullerenes. In fig (1-d) the entire spectrum is dominated by $C^+_{70}$ along with its fragments and the preceding ones of which it itself may be a fragment. $C^+_{60}$ is on the shoulder of the main peak whereas, $C^+_{50}$ and other clusters are present but much less intense than the main peak.

Figure (2) has four recoil spectra of $Xe^+$-C at 100 keV ion energy. The spectra are consecutively collected after every 30 minutes of bombardment. The beam incidence and recoil angles $\alpha$ and $\theta_r$, respectively are same as of fig (1). Figure 2(a) shows the snapshot of the target surface constituents after about 100 sec of bombardment at the ion arrival rate of $2x10^{+13}$ ions $s^{-1}$. The spectrum is divided approximately evenly between the $C^+_m < C^+_{36}$, centered around $C^+_{60}$ and $C^+_m > C^+_{70}$ clusters regimes. The next two spectra (2-b) and (2-c) clearly show the main contributions coming from fragments between $C^+_{82}$ and $C^+_{50}$. The cluster evolution continues in the next spectrum fig. 2(d) and is now dominated by $C^+_{70}$.

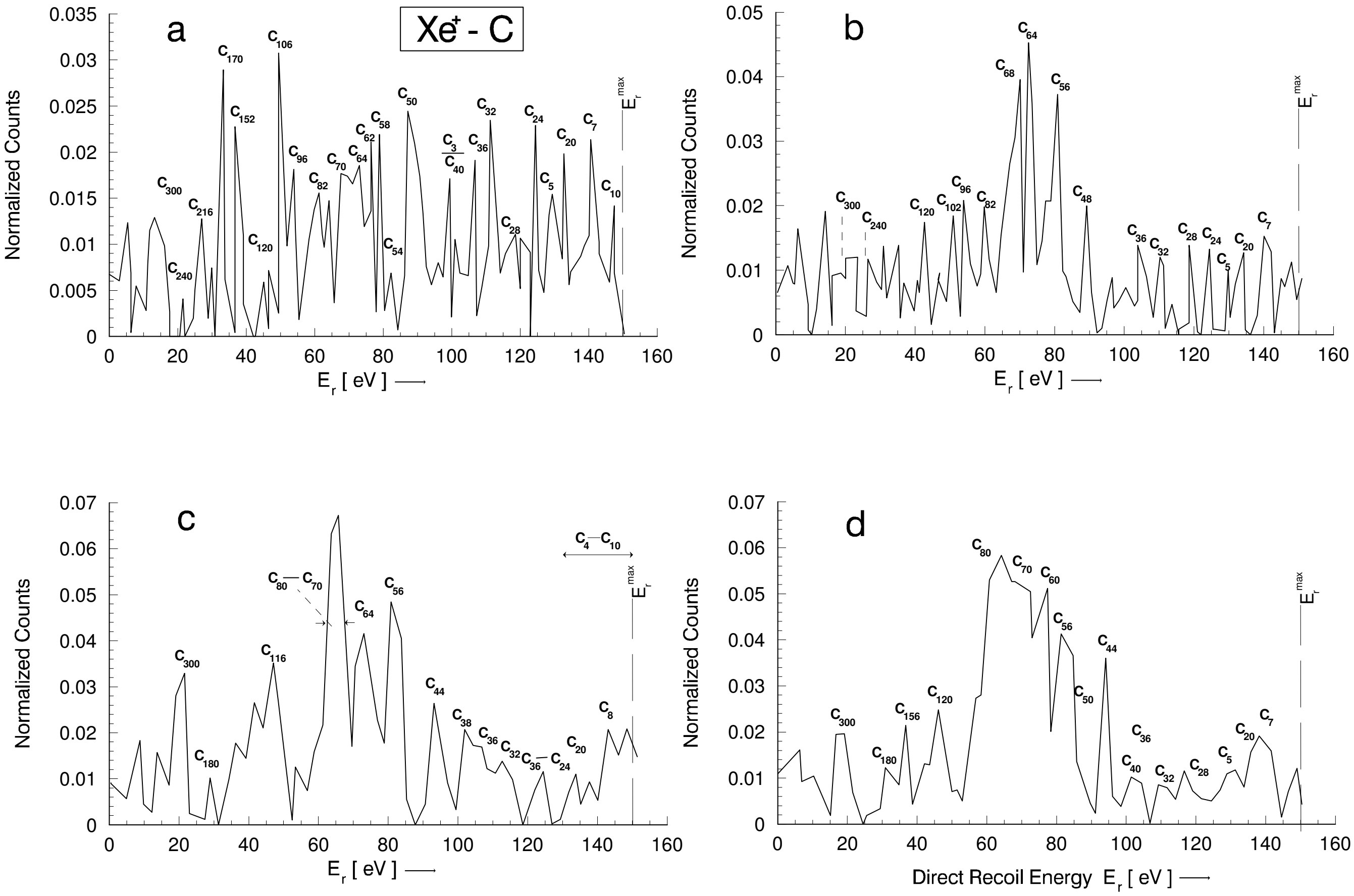

**Figure (2)**. DR energy spectra from 100 keV $Xe^+$ bombarded amorphous graphite is presented in the sequence of cumulatively increasing ion flux. The transition from an almost evenly distributed cluster spectrum in Fig (2a) to a gradual evolution of $C_m$ with $36<m\leq70$ regime is visible in fig.(2b), (2c) and more clearly in fig. (2d).

Figure (3-a) and (3-b) are plotted for the total accumulated counts in the three groups of clusters for $Kr^+$-C and $Xe^+$-C, respectively. The persistent trend of evolution of heavier fullerene structures with increasing fluence is obvious from the figure. The $Kr^+$-C results indicate that $C_{50}^{+}$- $C_{70}^{+}$ group gradually outnumbers other species while in $Xe^+$-C the cluster with m in the range > 50 up to 100 are all being produced with increasing ionic flux.

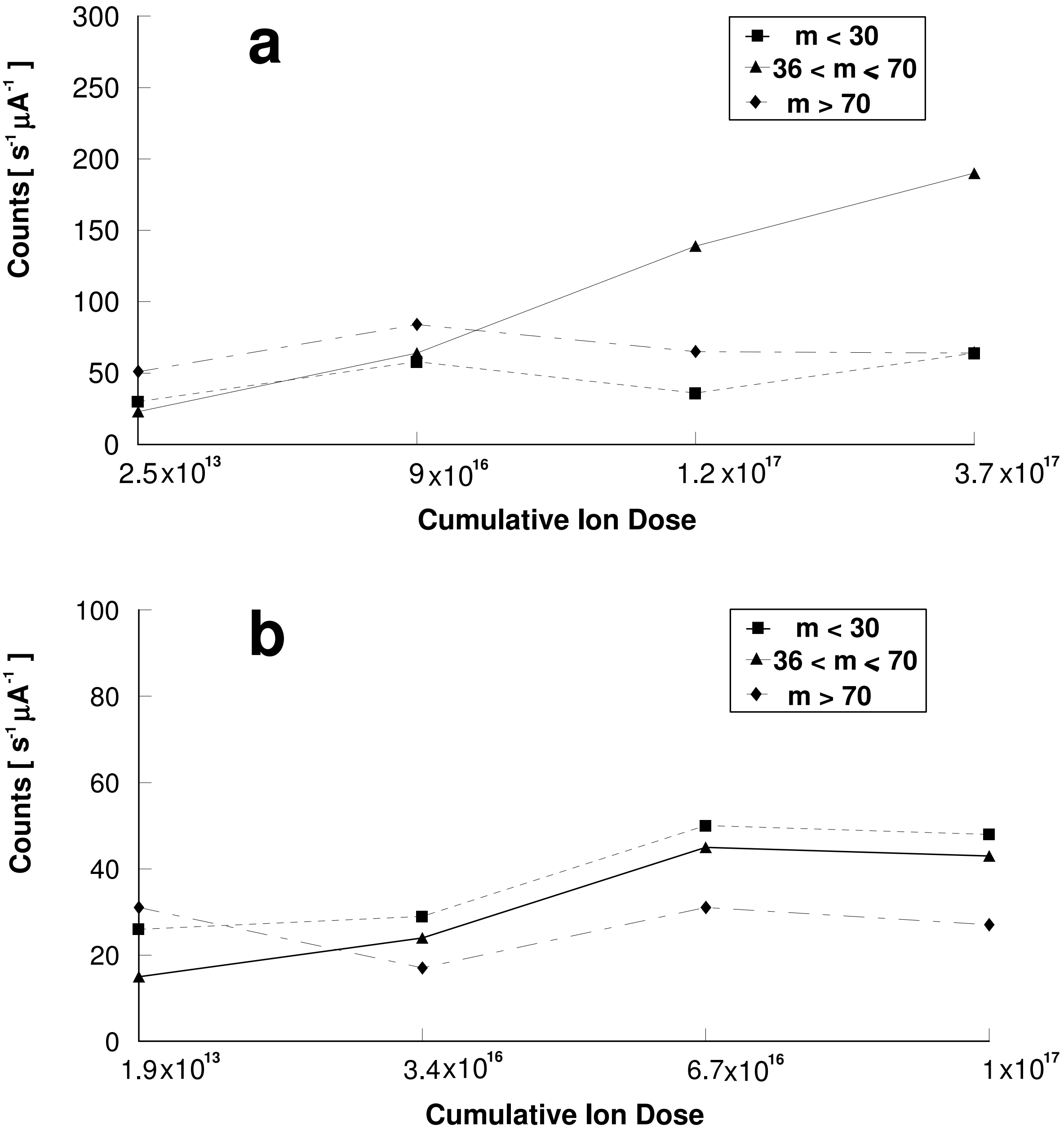

**Figure (3a)** $Kr^+$-C data from fig.(1) is computed for the total counts in the 3 cluster regimes m<36, 36<m≤70 and m>70 as a function of cumulative ion dose. The enhancement for 36<m≤70 regime is obvious. Fig (3b) shows the same data for $Xe^+$-C from fig.(2). Here all of the three regimes are enhanced with increasing dose. However, the net effect is to produce heavier clusters i.e. fullerenes ≥ $C_{50}$. The net effect is to selectively produce the main group of fullerenes with 36<m≤96.

The energy spectra reveal linear and chain complexes as well as fullerenes (m ≥ 20). The results clearly identify peaks corresponding to various clusters and their respective fragments which follow thee ion induced fragmentation processes. The presence of charged fullerenes in the direct recoil spectra implies their pre-existence prior to the primary knock-on collision. This information may help us to understand the structure and composition of the heavily irradiated 'amorphous graphite' surface and the mechanisms of direct recoiling of carbon clusters.

Heavy ion induced physical and chemical processes in graphite have been seen to lead to cluster formation as well as fragmentation. Our data suggest that the heavy ion irradiation has three essential features that make it unusual for the study of carbon clusters especially for fullerene formation.

(1) Almost all existing bonds between carbon atoms along the ionic path are broken on the time scale ~ $10^{-15}$ sec [18]. The primary, secondary and higher knock-ons being in highly excited/ionized states continue to produce energetically the most favorable carbon clusters. A cylindrical region with length equal to the respective ion range and radius of few Å is created by each incident ion within the irradiated surface. The ionic range determines the length of the cylinder while the radius depends upon the electronic stopping. The high energy of ions seems to provide the necessary environment for the evolution of the energetically the most favorable cluster population.

(2) A gradually expanding cylinder of increasing number density $\propto {E_r}^{-2}$ is generated which is dependent on nuclear stopping by the constituents of the energy dissipating medium. Time scales typical of the cooling down of the resulting intense collision cascades is ~ $10^{-13}$ -$10^{-12}$ sec.

(3) The Direct Recoils that originate in binary collisions between incident ions and the surface constituents act as snapshots of the surface that carry with them information characterizing the dynamics of cluster formation as well as fragmentation under ionic bombardment.

## Table 1

$Kr^+$ - C

| | | m > 70 | 36 < m ≤ 70 | m < 36 |
|---|---|---|---|---|
| **Fig (1)** | **a** | 0.29 | 0.22 | 0.49 |
| | **c** | 0.28 | 0.31 | 0.41 |
| | **d** | 0.15 | 0.58 | 0.27 |
| | **e** | 0.20 | 0.60 | 0.20 |

$Xe^+$ - C

| | | m > 70 | 36 < m ≤ 70 | m < 36 |
|---|---|---|---|---|
| **Fig (2)** | **a** | 0.37 | 0.21 | 0.43 |
| | **b** | 0.42 | 0.34 | 0.24 |
| | **c** | 0.43 | 0.30 | 0.27 |
| | **d** | 0.40 | 0.36 | 0.23 |

Table 1 summarizes the relative contributions of various carbon clusters $C_m^+$ in 3 main groups; (i) the largest closed structures with m > 70, (ii) the main group centered in the regime 36< m≤ 70 and (iii) the smaller fullerenes and multimers with m ≤ 36. The evolution of closed cages becomes statistically significant with $Kr^+$ irradiation.